\documentstyle[11pt]{article}

\begin{document}

\title{Targeting and control in simple dynamical systems}

\author{Girish Nathan\\\emph{Department of Physics and Astronomy}\\\emph{Rice University,
Houston, Texas 77250}}

\maketitle \vspace{1cm} \begin{abstract} A simple algorithm is described 
to target any desired operation point for simple
one-dimensional and two-dimensional dynamical systems. What makes the algorithm unique is the
fact that it targets any desired point, not merely a stable/unstable fixed point of the
dynamics. It is shown how the method may be applied to the logistic map and the H\'enon map.
Generalisations to the case of n-dimensional dynamical systems are discussed.

PACS number(s) : 05.45.-a,05.45.Gg
\maketitle
\end{abstract}

\section{Introduction} The idea of chaos control using only periodic
orbits was introduced by Ott,Grebogi and Yorke in their seminal paper[1].
Ever since, the ideas presented in that paper and subsequent modifications
of the algorithm [2-3]have been used to control chaotic dynamics in a
variety of experimental systems[4-6]. There have also been advances in
using noise to control the dynamics of systems, in particular by
Parmananda and co-workers[7], and by Ditto et al[8].

This paper seeks to answer the following question : Is it possible, using a targeting
algorithm, to (perturbatively) achieve any desired operation point for a dynamical system?
In particular, the focus is on points that are not necessarily
stable/unstable fixed points of the underlying
dynamics. It will be shown that the method to be described works very well for the 
logistic and the H\'enon map. 

The paper is organized as follows : In section 2, we introduce the basic
algorithm to be
employed to control the dynamics. We illustrate what a working model of
the
algorithm might look like, and try to motivate the reasons for its success. We
consider a general m-dimensional dynamical system and show how the algorithm works in
this case. Section
3 seeks to apply the algorithm to the logistic map. It is conclusively shown (both
theoretically and using numerical simulations) that the method works very
well for the logistic map.
In section 4  the algorithm is applied to the H\'enon map
-- it is again demonstrated that the algorithm is very effective and
rather fast in converging to the desired target point. In Section 5, we
show how the algorithm can be modified easily to include a stepwise
control, thus ensuring a gradual rate of convergence to the target point.
We end the paper with some conclusions.

\section{The Algorithm}

In this section, we describe in detail the targeting algorithm discussed
in the Introduction. For systems whose control parameter can be
varied in time, similar control mechanisms have been investigated by a
number of authors [9-10].The principal assumption throughout this paper is
that we are dealing with (a) systems where the control parameter cannot be
varied at all, (b) systems where it is necessary to sit at a fixed value
of
the control parameter and yet be able to work at a desired operation
point. The algorithm therefore assumes that the control parameter is fixed
throughout the duration of the dynamics of the system.

Consider a general $m$-dimensional dynamical system described by m coupled difference
equations :

\begin{equation}
x^{(n+1)}_{i}=f_{i}(x^{(n)}_{1},x^{(n)}_{2},\ldots,x^{(n)}_{m};\mu_{1},\mu_{2},\ldots,\mu_{m})
\end{equation}

Here, $n$ is the discrete time index and the variable $i$ runs from 1 to
m. The control
parameters are given by $\mu_{1}, \ldots ,\mu_{m}$  and $x^{(n)}_{i}$ are
the values of the
dynamical variables that describe the system ($x_{i}$) at time $n$. It is assumed
that the
dynamics of the system at time $n+1$ depend only on the values of the various dynamical
variables, $x_{i}$ at time $n$.

For the purposes of this discussion, as alluded to earlier, the $\mu_{i}$'s are fixed. Let
us say that a desired target state of the dynamical system (not necessarily a
stable/unstable fixed point of the dynamics) is $\vec{T}\equiv
(T_{1},T_{2},\ldots,T_{m})$. Given this desired target, the targeting algorithm works as
follows. We define a "distance" of the $i^{th}$ dynamical
variable,$x^{(n)}_{i}$ from its target
value $T_{i}$ by :-

\begin{equation}
d^{(n)}_{i}=\mid x^{(n)}_{i}-T_{i} \mid
\end{equation}

The new value of the dynamical variables describing the system are then given by the algorithm
as :-
\begin{equation}
X^{(n+1)}_{j}=x^{(n+1)}_{j} \prod_{i=1}^{m} d^{(n)}_{i}+T_{j}, \forall
j=1,\ldots,m
\end{equation}

We make the assignment $x^{(n+1)}_{j}=X^{(n+1)}_{j}$ firstly. Then the map
is iterated according to these modified values of the dynamical variables
and the procedure in Eqs.(2)-(3) is repeated. Notice how the algorithm works now. The
product term in (3) is zero when ANY $d^{n}_{j}$ vanishes. Since this
happens when $x^{n}_{j}=T_{j}$, it is clearly seen that the dynamics of
the modified system is such that it rapidly converges to the target point
once any one dynamical variable has reached its target value. In fact, it
can be seen from the above steps that it takes the system only one further
iteration to reach the desired target point, $\vec{T}$. Though the
algorithm appears
"uncontrolled" here, it will be demonstrated in Section 5 that this is not
so. A stepwise control is implemented for both the logistic and the
H\'enon map. In sections 3 and 4, we also show that the target point is
stable. We use numerical simulations to show the robustness of the
algorithm to noise.

 \section{The Logistic Map}
In this section, we apply the targeting algorithm introduced in the
previous section to the logistic map. In this section, since the system 
is a one-dimensional discrete dynamical system, the time index appears as
a
subscript only. The logistic map is defined by the difference equation :

\begin{equation} 
x_{n+1}=\mu x_{n} (1-x_{n})
\end{equation}

Here, $0 \leq \mu \leq 4$ and $0 \leq x_{n} \leq 1$. Let us call the
target point, $t$. Applying the algorithm to the logistic map then yields
the "modified logistic map",

\begin{equation}
x_{n+1}=\mu x_{n} (1-x_{n}) \mid x_{n}-t \mid+t
\end{equation}

As is seen, $t$ is a fixed point of the map.
Let us analyse the stability of this point. To do this, we
compute the derivative, $D=\frac{d x_{n+1}} {d x_{n}}$ at $t$. We
obtain $D=D_{1}+D_{2}+D_{3}$, where $D_{1}$,$D_{2}$, and $D_{3}$ are given
by :
\begin{equation}
D_{1}=\mu (1-x_{n})\mid x_{n}-t \mid
\end{equation}

\begin{equation}
D_{2}=-\mu x_{n} \mid x_{n}-t \mid
\end{equation}

\begin{equation}
D_{3}=\pm \mu x_{n} (1-x_{n})
\end{equation}

Combining all these expressions and evaluating the derivative at $t$, we
see that $D(t)=\pm \mu t (1-t)$. From the properties of the logistic map,
$\mid D(t) \mid \leq 1$ (the equality being reached only when $t=0.5$ and
$\mu=4$. Therefore, we see that
the new desired target point is indeed stable. The algorithm thus ensures
stability of the fixed point. A couple of points are in order here -- 
it is seen that applying the algorithm to the system does
not always ensure that the iterates are in the range $[0,1]$. This,
therefore, has to be enforced during the simulations. 

Having analysed the stability of the target point, we now show some
examples where the algorithm has been applied successfully. Figure 1 shows
the plot of the iterates versus the discrete time index,$n$, for the
uncontrolled logistic map for two values of the control parameter, $\mu$.
The corresponding plots in Figure 2 show the logistic map with the
targeting algorithm applied. Notice how quick the convergence to the
target point is in both cases. We have observed that the convergence is
typically that good. Though only two representative values of $\mu$ have been chosen here, the
algorithm has been tested over a large number of control parameter values and is found to work
just as well in all cases.

To look at how noise affects the ability of the algorithm to target, we
kicked the logistic map iterates randomly for the above two values of the
control parameter. The noise used to kick the iterates is uniform in $[0,1]$. Shown in Figure
3 are
plots of the iterates
versus the discrete time index,$n$. We see that the noise only momentarily
disturbs the targeting ability, and that the algorithm is indeed robust
against noise. This is a good feature, since noise is almost unavoidable
in experiments, and the algorithm would not be applicable to any real
experiments if it were sensitive to noise. 

\section{The H\'{e}non map}
 
In this section, we apply the algorithm to the H\'enon map and demonstrate
that targeting works very well here as well. 
The robustness of the algorithm to noise
is also investigated and it is shown that a conclusion similar to that
found for the logistic map holds. 

The H\'enon map is the simplest extension of the logistic map to the case
of 2 dimensions. In this section as in the previous one, the time index
appears as a subscript throughout. The H\'enon map is defined by the
following set of coupled difference equations:-
\begin{eqnarray}
x_{n+1} &= & 1-\mu x^{2}_{n}+\alpha y_{n} \nonumber \\
y_{n+1} &= & x_{n}
\end{eqnarray}

The map is invertible as long as $\alpha \neq 0$. Let the desired
target point in this case be $(x_{t},y_{t})$. Applying the algorithm
to the map yields the following "modified H\'enon map" :

\begin{equation}
x_{n+1}=(1-\mu x^{2}_{n}+\alpha x_{n}) M+x_{t}
\end{equation}
\begin{equation}
y_{n+1}=x_{n} M+y_{t}
\end{equation}

where $M=\mid x_{n}-x_{t} \mid \mid y_{n}-y_{t} \mid$.

To look at the stability of the target point $(x_{t},y_{t})$, we need to
examine the Jacobian of the system at that point. To evaluate the
derivatives of $x_{n+1}$ and $y_{n+1}$ at ($x_{t},y_{t}$), we use the fact
that $\mid z \mid = z \theta(z) - z \theta(-z)$, where $z=x_{n}-x_{t}$ or
$y_{n}-y_{t}$, and then take derivatives. Here we find something
remarkable. At the target
point, the Jacobian of the system is 0. Every term in the Jacobian is 0,
and therefore the eigenvalues of the Jacobian matrix are 0 too. From the
theory of fixed 
points, this implies a superstable fixed point. Therefore, the targeting
algorithm does indeed ensure that the desired target point is stable in
this case as well. 

Figures 4 shows the iterates of the uncontrolled map for two sets of
parameters, while Figure 5 shows the map for the same two sets of
parameters when the targeting algorithm is applied. We again see that the
number of
iterations taken for the system to settle down at the target point is very small. Similar
experiments have been performed over a fairly wide range of parameter values with similar
results. Figure 6 shows the effect of noise on the algorithm. It is again
seen clearly that the algorithm is robust to noise and that the system settles down from any
random perturbations very quickly to attain the desired target point. 

It is important that the initial point be in the basin of attraction for
the given parameter set. Numerical experiments were done where this was
not the case, and the dynamical system is quickly attracted to the point
at $\infty$ for almost any targeting point. 

\section{Stepwise Control Using The Algorithm}

In the previous sections as seen in the figures the approach to the target
point is not gradual. However, it is not hard to envisage applications
where it might be crucial to have a stepwise control whereby the target
point could be approached in a smooth fashion. In this section, it is
shown that the algorithm can be implemented in a way as to allow for a
gradual approach to the target point. 

To achieve a stepwise control, we break up the "distance" of the first
iterate from the target point into a large number of steps (for the
simulations shown, the number of steps is 1000). After that, the
targeting algorithm is just run so that each step is regarded as a
temporary target point, so that the system moves from one temporary target
point to another until it finally reaches the desired target point. By
choosing the steps to be small, one can therefore approach the target in a
controlled fashion. 

Figures 7 and 8 demonstrate this idea for the logistic and the H\'enon
map, respectively.

\section{Conclusions}

In conclusion, we have demonstrated a simple algorithm to target any
desired operation point for simple one and two dimensional systems.
In fact, as mentioned in Section 2, the algorithm can be easily
generalised
to a m-dimensional dynamical system. The advantages of the algorithm are
the fact that it can target any point (not necessarily a fixed point of
the system), it is very fast, and that it is fairly robust against noise
and random perturbations. The disadvantages are that one does need to know
the functional form of the equations describing the dynamical system, and that unlike some
other control algorithms, the control has to be applied during the full duration of the
dynamics of the system. Work is ongoing to see how this limitation can be
relaxed. We are also working on trying to implement (atleast theoretically) a simple
electrical circuit that would achieve the objective. The results of this investigation will
be published at a later date. 

\section{Acknowledgements}
Discussions with Gemunu Gunaratne are gratefully acknowledged.

\section{Captions}

\textbf{Figure 1a} : The uncontrolled logistic map for $\mu=1.05$. The iterates are plotted 
against the discrete time index $n$.

\textbf{Figure 1b} : The uncontrolled logistic map for $\mu=3.7$. The iterates are plotted
against the discrete time index $n$.

\textbf{Figure 2a} : The controlled logistic map for $\mu=1.05$, the target points being at
$t=0.75$ and $t=0.42$.

\textbf{Figure 2b} : The controlled logistic map for $\mu=3.7$, the target points being at
$t=0.005$ and $t=0.95$.

\textbf{Figure 3a} : The robustness of the algorithm to noise - Random kicks applied once
every 100 iterations to the logistic map for $\mu=1.05$. The dotted line denotes iterations
for the target point $t=0.42$, and the solid line denotes iterations for the target point
$t=0.75$.

\textbf{Figure 3b} : Random kicks applied once every 100 iterations to the logistic map for
$\mu=3.7$. The dotted line denotes iterations for the target point $t=0.75$ and the solid
line
denotes iterations for the target point $t=0.42$.

\textbf{Figure 4a} : The uncontrolled H\'enon map for parameter values,
$\mu=1.1$ and $\alpha=0.3$, with $y_{n}$ plotted against $x_{n}$.

\textbf{Figure 4b} : The uncontrolled H\'enon map for parameter values,
$\mu=1.1$ and $\alpha=-0.3$, with $y_{n}$ plotted against $x_{n}$.

\textbf{Figure 5a} : The controlled H\'enon map for $\mu=1.1$ and $\alpha=0.3$ with $y_{n}$
plotted against $x_{n}$ - the target
.The solid line denotes initial point at $(-0.005,-0.5)$ and
target point at $(-1,-1)$. The dotted line denotes initial point at
$(-0.1,-0.1)$ and target point at $(0.9,0.9)$. The small dashed line
denotes initial point at $(0.7,-0.4)$ and target point at $(0,0.5)$. The
large dashed line denotes initial point at $(0.95,0.5)$ and target point
at $(-1,-1)$.

\textbf{Figure 5b} : The controlled H\'enon map for $\mu=1.1$ and $\alpha=-0.3$
with $y_{n}$ plotted against $x_{n}$. The lines, initial and target points are exactly the
same
as
for Figure 5a.

\textbf{Figure 6a} : Robustness of the algorithm to noise - demonstrated for the H\'enon
map
for parameter values $\mu=1.1$ and $\alpha=0.3$. The initial point is $(-0.5,0.5)$, the
target
point
is $(0.2,1.0)$ and the kicks are applied once every 100 iterations.

\textbf{Figure 6b} : Robustness of the algorithm to noise - demonstrated for the H\'enon map
for parameter values $\mu=1.1$ and $\alpha=-0.3$.The initial point is $(-0.5,0.5)$, the
target
point
is $(0.2,1.0)$ and the kicks are applied once every 100 iterations.

\textbf{Figure 7} : Stepwise control of the logistic map for $\mu=1.05$ - the initial point
is $0.005$, the target point is
$t=0.95$ and the number of steps is 1000. The achieved target point is plotted versus the
iteration index, $n$.

\textbf{Figure 8} : Stepwise control of the H\'enon map for parameter values $\mu=1.1$ and
$\alpha=0.3$ - the initial point is $(0.5,0.5)$, the target point is
$(-1,-1)$ and the number of steps is 1000. The achieved target point is plotted versus
iteration index, $n$.

 \end{document}